\documentclass[journal,10pt]{IEEEtran}

\IEEEoverridecommandlockouts

\usepackage[cmex10]{amsmath}
\usepackage[dvips]{graphicx}
\usepackage{cite}
\usepackage{amssymb}
\usepackage[eulergreek]{sansmath}
\usepackage{color}
\usepackage{balance}
\usepackage{graphicx,floatrow,subfig}
\graphicspath{{Simulation-Figures/}}

\begin{document}

\title{mmWave Doubly-Massive-MIMO Communications Enhanced with an Intelligent Reflecting Surface}
\author{Dian-Wu Yue, Ha H. Nguyen, and Yu Sun
\thanks{Dian-Wu Yue is with the College of Information Science and
Technology, Dalian Maritime University, Dalian, Liaoning 116026,
China (e-mail: dwyue@dlmu.edu.cn), and also with the Department of Electrical and Computer Engineering, University of Saskatchewan, 57 Campus Drive, Saskatoon, SK, Canada S7N 5A9.}
\thanks{Ha H. Nguyen is with the Department of Electrical and Computer Engineering, University of Saskatchewan, 57 Campus Drive, Saskatoon, SK, Canada S7N 5A9 (e-mail: ha.nguyen@usask.ca).}
\thanks{Yu Sun is with the College of Information Science and
Technology, Dalian Maritime University, Dalian, Liaoning 116026, China (e-mail:suny@dlmu.edu.cn).}
}


\newcommand{\be}{\begin{equation}}
\newcommand{\ee}{\end{equation}}
\newcommand{\bee}{\begin{eqnarray}}
\newcommand{\eee}{\end{eqnarray}}
\newcommand{\nnb}{\nonumber}

\newcommand{\mo}{\mathbf{0}}
\newcommand{\mA}{\mathbf{A}}
\newcommand{\mB}{\mathbf{B}}
\newcommand{\mG}{\mathbf{G}}
\newcommand{\mH}{\mathbf{H}}
\newcommand{\mI}{\mathbf{I}}
\newcommand{\mR}{\mathbf{R}}
\newcommand{\mY}{\mathbf{Y}}
\newcommand{\mZ}{\mathbf{Z}}
\newcommand{\mD}{\mathbf{D}}
\newcommand{\mW}{\mathbf{W}}
\newcommand{\mF}{\mathbf{F}}
\newcommand{\mP}{\mathbf{P}}
\newcommand{\mQ}{\mathbf{Q}}
\newcommand{\mU}{\mathbf{U}}
\newcommand{\mV}{\mathbf{V}}
\newcommand{\mPsi}{\mathbf{\Psi}}
\newcommand{\mSigma}{\mathbf{\Sigma}}

\newcommand{\my}{\mathbf{y}}
\newcommand{\mx}{\mathbf{x}}
\newcommand{\mz}{\mathbf{z}}

\newcommand{\mr}{\mathbf{r}}
\newcommand{\mt}{\mathbf{t}}
\newcommand{\mb}{\mathbf{b}}
\newcommand{\ma}{\mathbf{a}}

\newcommand{\mh}{\mathbf{h}}
\newcommand{\mw}{\mathbf{w}}
\newcommand{\mg}{\mathbf{g}}
\newcommand{\mf}{\mathbf{f}}
\newcommand{\mn}{\mathbf{n}}

\newcommand{\mv}{\mathbf{v}}
\newcommand{\ms}{\mathbf{s}}
\newcommand{\mmu}{\mathbf{u}}

\newcommand{\lf}{\left}
\newcommand{\ri}{\right}

\newtheorem{Remark}{Remark}
\newtheorem{Lemma}{Lemma}
\newtheorem{Theorem}{Theorem}
\newtheorem{Corollary}{Corollary}
\newtheorem{Proposition}{Proposition}
\newtheorem{Example}{Example}
\newtheorem{Definition}{Definition}


\maketitle

\begin{abstract}
As a means to control wireless propagation environments, the use of emerging and novel intelligent reflecting surfaces (IRS) is envisioned to enhance and broaden many applications in future wireless networks. This paper is concerned with a point-to-point IRS-assisted millimeter-wave (mmWave) system in which the IRS consists of multiple subsurfaces, each having the same number of passive reflecting elements, whereas both the transmitter and receiver are equipped with massive antenna arrays. Under the scenario of having very large numbers of antennas at both transmit and receive ends, the achievable rate of the system is derived. Furthermore, with the objective of maximizing the achievable rate, the paper presents optimal solutions of power allocation, precoding/combining, and IRS's phase shifts. Then it is shown that when the number of reflecting elements at each subsurface is very large, the number of favorable and controllable propagation paths provided by the IRS is simply equal to the number of subsurfaces while the received signal-to-noise ratio corresponding to each of the favorable paths increases quadratically with the number of reflecting elements. In addition, the problem of minimizing the transmit power subject to the rate constraint is analyzed for the scenario without direct paths in the pure LOS propagation. Finally, numerical results are provided to corroborate the obtained analysis.
\end{abstract}


\section{Introduction}
As a key enabling technology for 5G mobile communication systems, millimeter-wave (mmWave) communication has recently gained considerable attentions in both research community and industry \cite{Andrews}. In mmWave communications, to compensate for the very high propagation loss, the use of compact massive antenna arrays is quite natural and attractive \cite{Heath}, \cite{Molisch}. Since a very large antenna array can be realized in a very small volume, it is practical to mount large numbers of antennas at both the transmit and receive terminals. Such a MIMO system is called a mmWave doubly-massive MIMO system \cite{Buzzi1}, \cite{Buzzi2}, \cite{Yue1}. In this paper, we shall consider a mmWave doubly-massive MIMO system that is enhanced by making use of an intelligent reflecting surface (IRS).

IRS, also known as \emph{reconfigurable reflecting surface}, \emph{large intelligent surface}, and \emph{software-controlled metasurface}, is a recently emerging novel hardware technology that can extend signal coverage, reduce energy consumption and enjoy low-cost implementation \cite{Wu1,Basar1,Liaskos,Bjornson}. Different from cooperative relaying and backscatter communications, an IRS consists of a large number of small, passive, and low-cost reflecting elements, which only reflect the incident signal with an adjustable phase shift without requiring a dedicated energy source for RF processing, encoding/decoding, and retransmission. Because of their attractive advantages, IRS has been rapidly introduced into various wireless communication systems \cite{Wu2}--\cite{Guan}.

There is a large body of works focusing on performance analysis and optimization design of IRS-aided MIMO systems in traditional microwave bands \cite{Wu1}--\cite{Guan}. In particular, for a single-user scenario,  it is shown in \cite{Wu2} that the received signal power of the IRS-aided system increases quadratically with the number of reflecting elements. For the multiuser scenario, it is shown in \cite{Nadeem} that the IRS-aided system can offer massive MIMO-like gains with a much smaller number of active antennas. To date, there are a few works that investigate IRS-aided mmWave MIMO systems for both single-user and multi-user scenarios, and with one or multiple IRSs. These works were concerned with rate optimization \cite{Cao,Guo,Perovic}, joint active and passive beamforming design \cite{Wang1,Wang2}, joint power allocation and beamforming design \cite{Xiu}, hybrid beamforming design \cite{Pradhan,Ying}, and channel estimation \cite{Wang3}.

For a mmWave communication system, the severe path loss and high directivity make it vulnerable to blockage events, which can be frequent in indoor and dense urban environments \cite{Wang1}. Moreover, due to the multipath sparsity of mmWave signal propagation, the potential of spatial multiplexing is limited \cite{Basar1,Yue2}. To address the blockage issue and enhance spatial multiplexing, this paper considers an IRS-assisted mmWave doubly-massive MIMO system, which, to the best of our knowledge, has not been sufficiently studied in literature \cite{Yue3}. In particular, this paper examines the spatial multiplexing potential provided by the IRS consisting of multiple subsurfaces for the very high date rate requirement in future wireless communications. The contributions of this paper are summarized as follows:
\begin{itemize}
  \item With the clustered statistical channel model, we consider a point-to-point mmWave doubly-massive MIMO system assisted by an IRS consisting of multiple subsurfaces and derive an expression of the achievable rate of the system in the case of having very large numbers of antennas at both transmit and receive ends.
  \item Under the objective of maximizing the achievable rate of the mmWave doubly-massive system, we present optimal solutions of power allocation, precoding/combining, and IRS's phase shifts, and show that these optimal solutions can be realized independently.
  \item When the number of reflecting elements at each subsurface is also very large, we show that the number of favorable propagation paths provided by the IRS equals the number of subsurfaces and the received signal-to-noise ratio (SNR) corresponding to each of the favorable paths increases quadratically with the number of reflecting elements.
  \item For the scenario without direct signal paths in the pure LOS propagation, we derive the minimum transmit power with equal power allocation (EPA) and present an optimal solution of the number of subsurfaces when the total number of all reflecting elements of the IRS is given.
\end{itemize}

The rest of the paper is organized as follows. In Section II, the IRS-assisted system and channel model are described. In Section III, the asymptotic achievable rate is analyzed and optimal solutions of power allocation, precoding/combining, and IRS's phase shifts are presented. In Section VI, the scenario without direct paths is considered and the problem of minimizing the transmit power with equal power allocation is analyzed. Numerical results are provided in Section V, followed by concluding remarks in Section VI.

Throughout this paper, the following notations are used. Boldface upper and lower case letters denote matrices and column vectors, respectively. The superscripts $(\cdot)^T$ and $(\cdot)^H$ stand for transpose and conjugate-transpose, respectively. $\mathrm{diag}\{a_1,a_2,\ldots,a_N\}$ stands for a diagonal matrix with diagonal elements $\{a_1,a_2,\ldots,a_N\}$. The expectation operator is denoted by $\mathbb{E}(\cdot)$. $\mI_M$ is the $M \times M$ identity matrix. Finally, $\mathcal{N}(0, \sigma^2)$ or $\mathcal{CN}(0, \sigma^2)$ denotes a Gaussian or circularly symmetric complex Gaussian random variable with zero mean and variance $\sigma^2$.

\section{System Description}

\begin{figure}[t]
\centering
\includegraphics[width=3.3in]{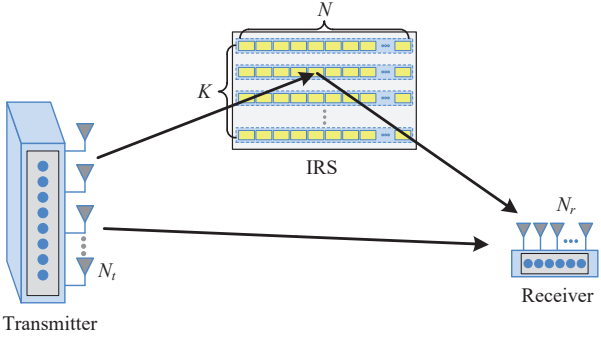}
\caption{IRS-assisted mmWave massive MIMO single-user system}
\label{SYS1}
\end{figure}

The IRS-assisted mmWave doubly-massive MIMO single-user (point-to-point) system under consideration is illustrated in Fig.~\ref{SYS1}, where an IRS is used to assist the transmission of multiple data streams from the transmitter to the receiver. The IRS consists of $K$ subsurfaces, each having $N$ reflecting elements arranged in a uniform linear array (ULA). The transmitter (source) is equipped with a large $N_t$-element ULA, while the receiver (destination) is equipped with a large $N_r$-element ULA.

Denote by $\mH_{\mathrm{TR}}$ the channel matrix between the transmitter and the receiver. In general, the matrix $\mH_{\mathrm{TR}}$ consists of a LOS matrix and a scattered matrix \cite{Yue1}, i.e., \be \mH_{\mathrm{TR}}=\sqrt{\bar{\eta}/g_0} \cdot \overline{\mH}_{\mathrm{TR}}+\sqrt{\tilde{\eta}/g_0} \cdot \widetilde{\mH}_{\mathrm{TR}}\ee where $g_0$ represents the large scale fading effect, which is assumed to be constant over many coherence-time intervals. When $g_0$ is large, the propagation loss will be high. In addition, $\bar{\eta}=\frac{\eta}{1+\eta}$, $\tilde{\eta}=\frac{1}{1+\eta}$, and $\eta \geq 0$ is the Ricean $K$-factor, which represents the relative strength of the LOS component. Note that if the LOS path between the transmitter and the receiver is blocked, then $\eta=0$ in the above model.

Since both of the antenna configurations at the transmit and receive ends are ULAs, the LOS matrix $\overline{\mH}_{\mathrm{TR}}$ can be written as  \be
  \overline{\mH}_{\mathrm{TR}}=\sqrt{N_rN_t}\ma_r(\phi_r^1)\ma_t^H(\theta_t^1). \ee
In the above equation, the vectors $\ma_r(\phi_r^1)$ and $\ma_t(\theta_t^1)$ are the normalized receive/transmit array response vectors at the corresponding angles of arrival and departure, $\phi_r^1$ and $\theta_t^1$, respectively. For an $M$-element ULA, the
array response vector is \be \ma(\phi)=\frac{1}{\sqrt{M}}\left[1,{\mathrm e}^{j2\pi\frac{d}{\lambda}\sin(\phi)},\ldots,{\mathrm e}^{j2\pi(M-1)\frac{d}{\lambda}\sin(\phi)}\right]^T \ee
where $\lambda$ is the wavelength of the carrier and $d$ is the inter-element spacing. As common in many references, it is assumed that $d=\frac{\lambda}{2}$.

Regarding the scattered matrix $\widetilde{\mH}_{\mathrm{TR}}$, limited-scattering mmWave fading is considered in this paper. Specifically, the clustered model (based on the extended Saleh-Valenzuela model) is used to characterize limited-scattering mmWave fading \cite{Ayach}. For simplicity of exposition, each scattering cluster is assumed to contribute a single propagation path. So $\widetilde{\mH}_{\mathrm{TR}}$ can be expressed as
\be \widetilde{\mH}_{\mathrm{TR}}=\sqrt{\frac{N_rN_t}{L}}\sum_{l=2}^{L}\alpha^{l}\ma_r(\phi^{l}_r)\ma_t^H(\theta^{l}_t), \ee
where $L$ is the number of total propagation paths (including the LOS path), $\alpha^{l}$ is the complex gain of the $l$th ray, and $\phi^{l}_r$ and $\theta^{l}_t$ are random azimuth angles of arrival and departure, respectively. Without loss of generality, the complex gains $\alpha^{l}$ are assumed to be $\mathcal{CN}(0, 1)$.

Throughout the paper, it is assumed that the space between adjacent subsurfaces is much larger than the wavelength so that the channels of different subsurfaces are spatially independent. For the $k$th subsurface, due to the ULA element configuration,  the channel matrix between the transmitter and the subsurface is also described as
\be \mH_{\mathrm{TI}}^{k}=\sqrt{\bar{\eta}^k_1/g_1^k} \cdot \overline{\mH}_{\mathrm{TI}}+\sqrt{\tilde{\eta}^k_1/g_1^k} \cdot \widetilde{\mH}_{\mathrm{TI}}, \ee
where
\be \overline{\mH}_{\mathrm{TI}}^k=\sqrt{N_t N}\ma_1(\phi_1^{k1})\ma_1^H(\theta_1^{k1}), \ee
and
\be \widetilde{\mH}_{\mathrm{TI}}^k=\sqrt{\frac{N_tN}{L^k_1}}\sum_{j=2}^{L^k_1}\alpha^{kj}_1\ma_1(\phi^{kj}_1)\ma_1^H(\theta^{kj}_1). \ee
When the channel is LOS dominated \cite{Brady},  $\eta^k_1$ can be assumed to be large enough. Similarly, the channel between the subsurface and the receiver is also modelled as
\be \mH_{\mathrm{IR}}^{k}=\sqrt{\bar{\eta}^k_2/g_2^k} \cdot \overline{\mH}_{\mathrm{IR}}+\sqrt{\tilde{\eta}^k_2/g_2^k} \cdot \widetilde{\mH}_{\mathrm{IR}}, \ee
where
\be \overline{\mH}_{\mathrm{IR}}^k=\sqrt{N_r N}\ma_2(\phi_2^{k1})\ma_2^H(\theta_2^{k1}), \ee
and
\be \widetilde{\mH}_{\mathrm{IR}}^k=\sqrt{\frac{N_rN}{L^k_2}}\sum_{i=2}^{L^k_2}\alpha^{ki}_1\ma_2(\phi^{ki}_2)\ma_2^H(\theta^{ki}_2). \ee
It should be pointed out that all of the complex gains, $\alpha^{kl}_b, \; l=2, 3, \ldots, L^k_b; \;  b=1, 2,$ are also assumed to be $\mathcal{CN}(0, 1)$.

Furthermore, each of the above mentioned large scale fading parameters, $g$,  can be described via a linear model of the following form \cite{Akdeniz}
\be g\; [\mathrm{dB}] = a(g) + b(g) \log_{10} (d(g)) + \chi(g), \ee
where $d(g)$ is the distance, $a(g)$ and $b(g)$ are linear model parameters and $\chi(g) \sim \mathcal{N}(0, \sigma^2(g))$ is a lognormal term accounting for variances in shadowing. Additionally, it can be assumed that both of the distances between the transmitter and the IRS and between the IRS and the receiver are much larger than the size of the IRS. For simplicity, we further assume that for any $k$, $a(g_1^k)=a_1$, $a(g_2^k)=a_2$, $b(g_1^k)=b_1$, $b(g_2^k)=b_2$, $d(g_1^k)=d_1$, $d(g_2^k)=d_2$, $\sigma^2(g_1^k)=\sigma^2_1$, $\sigma^2(g_2^k)=\sigma^2_2$. For brevity, we introduce the notation $a(g_0)=a_0$, $b(g_0)=b_0$, $d(g_0)=d_0$, and $\sigma^2(g_0)=\sigma^2_0$.

The IRS is intelligent in the sense that each of the reflecting elements can control the phase of its diffusely reflected signal. In particular, the reflection properties of the $k$th subsurface are determined by the following diagonal matrix \be \mV^k=\beta\cdot\mathrm{diag}\{\mathrm{e}^{-jv_1^k},\mathrm{e}^{-jv_2^k},\ldots,\mathrm{e}^{-jv_N^k} \}, \ee
where $\beta \in (0, 1]$ is a fixed amplitude reflection coefficient and $v_1^k, v_2^k,\ldots,v_N^k$ are the phase-shift variables that can be optimized by the IRS based on the known channel state information (CSI) and requirements of the system design. For simplicity, we assume that $\beta=1$ throughout this paper.
It follows that the overall channel matrix of the IRS-assisted mmWave MIMO system can be expressed as
\be \mathbb{H}=\sum_{k=1}^K\mH_{\mathrm{IR}}^{k}\mV^k\mH_{\mathrm{TI}}^{k}+\mH_{\mathrm{TR}}. \ee

Suppose that the  matrix  $\mathbb{H}$ has a rank of $r$. Then we can use the MIMO channel to transmit $N_s \leq r$ data streams.  The transmitter accepts as its input $N_s$ data streams and applies a $N_t \times N_s$  precoder, $\mW_t$. Then the transmitted signal vector can be written as
\be \mx = \mW_t\mP_t^{1/2}\ms, \ee
where  $\ms$ is the $N_s  \times  1$ symbol vector such that $\mathbb{E}[\ms\ms^H] = \mI_{N_s}$, and $\mP_t=\mathrm{diag}\{p_1,\ldots,p_{N_s}\}$ is a diagonal power allocation matrix with $\sum_{l=1}^{N_s}p_{l}=P$ and $p_l>0$. Thus $P$ represents the average total input power. Then the $N_r \times 1$ received signal vector is \be \label{my} \my =\mathbb{H}\mW_t\mP_t^{1/2}\ms + \mn,  \ee
where $\mn$ is a $N_r \times 1$ vector consisting of independent and identically distributed (i.i.d.) $\mathcal{CN}(0, \sigma^2_n)$ noise samples. Throughout this paper, all of the channel state information (CSI) involving $\{\mH_{\mathrm{TR}}, \mH_{\mathrm{TI}}^k, \mH_{\mathrm{IR}}^k\}$ is assumed known to both the transmitter and receiver \footnote{Channel estimation is a very important problem \cite{Wu1,Basar1}, which will be considered in our future work.}. Such an assumption allows us to focus on analyzing the absolute performance limit of the system. Let $\mW_r$ denotes the $N_r \times N_s $ combining matrix at the receiver. The processed signal for detection of the $N_s$ data streams is given by
\be  \mz =\mW_r^H\mathbb{H}\mW_t\mP_t^{1/2}\ms + \mW_r^H\mn.  \ee

We define $\mathcal{V}=\{v_n^k\}$ and $\mathcal{P}=\{p_{l}\}$. In the next section, our goal is to find the optimal $\mathcal{V}$, $\mW_t$, $\mW_r$, and $\mathcal{P}$ to maximize the system's achievable data rate and further evaluate its multiplexing capability.

\section{Analysis of Asymptotic Achievable Rate}
With perfect CSI, the optimal precoding/combining and power allocation for a point-to-point wireless system can be achieved by applying singular value decomposition (SVD) and performing waterfilling. Let $\mU$ and $\mQ$ denote the right and left singular matrices of the channel matrix $\mathbb{H}$. Then the SVD factorizes the channel matrix as $\mathbb{H}=\mU\mSigma\mQ^H$ where $\mSigma=\mathrm{diag}\{\lambda_1,\lambda_2,\ldots,\lambda_r, 0, \ldots,0\}$, $\lambda_i$ stands for the $i$th effective singular value. By setting the precoding and combining matrices as  $\mW_t=\mQ_{1:r}$ and $\mW_r=\mU_{1:r}$, the maximum achievable sum rate can be obtained and expressed as
\bee  R&=&\max_{\{p_l\}}\log_2\det\left(\mI_r+\frac{\mB^{-1}}{\sigma^2_n}\mW_r^H\mathbb{H}\mW_t\mP_t\mW_t^H\mathbb{H}^H\mW_r\right)\nnb\\
&=&\max_{\{p_l\}}\sum_{l=1}^r\log_2\left(1+p_l\lambda_l^2/\sigma^2_n\right), \eee
where $\mB=\mW_r^H\mW_r=\mI_r$. The optimal power allocation $\{p_l\}$ can be obtained based on the well-known waterfilling procedure \cite{Tse}.

For the direct propagation paths, we introduce the following notations for brevity:
 \be \nu_0=\sqrt{\frac{\bar{\eta}N_rN_t}{g_0}}, \;\text{and}\; \nu_l=\sqrt{\frac{\tilde{\eta}N_rN_t}{Lg_0}}\alpha^{l}, \;\; l\geq 2.    \ee
Now define $\xi^k_{ij}=\ma_2^H(\theta_2^{ki})\mV^k\ma_1(\phi_1^{kj})$. Furthermore, we introduce the following notations for the IRS propagation paths:
 \be \label{var} \vartheta^k_{ij}=\sqrt{\frac{N_tN_r N^2}{g_1^kg_2^k}}\xi^k_{ij}\beta^{kj}_1 \beta^{ki}_2, \ee
where  \be \beta^{k1}_b=\sqrt{\bar{\eta}^k_b},\;\text{and}\; \beta^{kl}_b=\sqrt{\tilde{\eta}^k_b/L^k_b}\alpha^{kl}_b, \;\;  l\geq 2,\;  b=1, 2.\ee

Then the channel matrix $\mathbb{H}$ can be rewritten as
\bee \label{HHH} \mathbb{H}&=&\sum_{l=1}^{L}\nu_l\ma_r(\phi^{l}_r)\ma_t^H(\theta^{l}_t) \nnb\\
&&+\sum_{k=1}^K\sum_{i=1}^{L^k_2}\sum_{j=1}^{L^k_1}\vartheta^k_{ij}\ma_2(\phi_2^{ki})\ma_1^H(\theta_1^{kj})
\eee
The expression (\ref{HHH}) implies that the underlying IRS-assisted mmWave MIMO channel can be considered as a traditional mmWave MIMO channel with $L_K$ propagation paths, where $L_K=L+\sum_{k=1}^K(L^k_1L^k_2)$. More precisely, the $L_K$ paths have complex gains $\{\vartheta^k_{ij}, \nu_l\}$, receive array response vectors $\{\ma_2(\phi_2^{ki}),\ma_r(\phi^{l}_r)\}$ and transmit response vectors $\{\ma_1(\theta_1^{kj}), \ma_t(\theta^{l}_t)\}$. Furthermore, by ordering all paths in a decreasing order of the absolute values of  $\{\vartheta^k_{ij}, \nu_l\}$ and redefining the variables for complex gains and various angles as $\{\tilde{\nu}_l\}$, $\{\tilde{\phi}^{l}_r\}$, $\{\tilde{\theta}^{l}_t\}$, the channel matrix can be reexpressed as
\be  \mathbb{H}=\sum_{l=1}^{L_K}\tilde{\nu}_l\ma_r(\tilde{\phi}^{l}_r)\ma_t^H(\tilde{\theta}^{l}_t),   \ee
where $|\tilde{\nu}_1|\geq |\tilde{\nu}_2|\geq \cdots \geq |\tilde{\nu}_{L_K}|$.

\begin{Proposition} \label{Pro1} In the limit of large $N_t$ and $N_r$, the rank of the channel matrix $\mathbb{H}$ is equal to $r=L_K$ and the system's achievable rate is given by
\be \label{RRR} R=\sum_{l=1}^{L_K}\log_2(1+p_l|\tilde{\nu}_l|^2/\sigma^2_n).    \ee
\end{Proposition}
{\em Proof:} One can rewrite $\mathbb{H}$ in the following form:
\be  \mathbb{H}=\mA_r\mD\mA_t^H,  \ee
where $\mD$ is a $L_K \times L_K$ diagonal matrix with $[\mD]_{ll}=\tilde{\nu}_l$, and $\mA_r$ and  $\mA_t$ are defined as follows:
\be  \mA_r=[\ma_r(\tilde{\phi}^{1}_r),\ma_r(\tilde{\phi}^{2}_r),\ldots,\ma_r(\tilde{\phi}^{L_K}_r)],  \ee
and
      \be  \mA_t=[\ma_t(\tilde{\theta}^{1}_t),\ma_t(\tilde{\theta}^{2}_t),\ldots,\ma_t(\tilde{\theta}^{L_K}_t)].  \ee
It follows from \cite{Yue2} that all $L_K$ vectors $\{\ma_r(\tilde{\phi}^{l}_r)\}$ are orthogonal to each other when $N_r \to \infty$. Likewise, all $L_K$ vectors $\{\ma_t(\tilde{\theta}^{l}_t)\}$ are orthogonal to each other when $N_t \to \infty$. Thus $\mA_r$ and  $\mA_t$  are asymptotically unitary matrices under the limit of large $N_t$ and $N_r$. Then the SVD of matrix $\mathbb{H}$ can be formed as
\be \label{SVD}\mathbb{H}=\mU\mSigma\mQ^H=[\mA_r|\mA_r^{\bot}]\mSigma [\tilde{\mA}_t|\tilde{\mA}_t^{\bot}]^H,  \ee
where $\mSigma$ is a diagonal matrix containing all singular values on its diagonal, i.e.,
 \be [\mSigma]_{ll}=\left\{\begin{array}{ll}
 |\tilde{\nu}^l|, & \mbox{for}\; 1 \leq  l\leq L_K\\
0, & \mbox{for}\; l>L_K
     \end{array}
     \right. \ee
and the submatrix $\tilde{\mA}_t$ is defined as
\be  \tilde{\mA}_t=[{\mathrm e}^{j\psi_1}\ma_t(\tilde{\theta}^{1}_t),\ldots,{\mathrm e}^{j\psi_{L_K}}\ma_t(\tilde{\theta}^{L_K}_t)],   \ee
where $\psi_l$ is the phase of complex gain $\tilde{\nu}_l$ corresponding to the $l$th path. Since there exist only $L_K$ effective singular values in the channel matrix $\mathbb{H}$,  the rank of the channel matrix $\mathbb{H}$ is equal to $r=L_K$. Furthermore, the optimal precoder and combiner are given by
\be \label{WWW} [\mW_t]_\mathrm{opt}=\tilde{\mA}_t,\;\; [\mW_r]_\mathrm{opt}=\mA_r.   \ee
Finally, it is easy to prove that (\ref{RRR}) holds.
\hfill $\square$

\begin{Remark} \label{Re1} Traditional, the fully-digital precoding/combining architecture for a mmWave massive MIMO system is expensive. However, (\ref{WWW}) implies that instead of the fully-digital architecture, a cost-efficient analog precoding/combining architecture can be applied in the underlying IRS-assisted mmWave massive MIMO system. Furthermore, the optimal precoding/combining matrix can be determined, provided that the angles of departure and arrival related to the transmit and receive terminals are obtained. In practice, the employment of hybrid precoding/combining (beamforming) instead of pure analog precoding/combining (beamforming) can lead to further improvement of the system performance \cite{Heath,Molisch}.
\end{Remark}

Given $\{\tilde{\nu}_l\}=\{\vartheta^k_{ij}, \nu_l\}$, the optimal power allocation can be obtained immediately once the parameter set $\{\vartheta^k_{ij}, \nu_l\}$ is known. Since $\{\nu_l\}$ is known, in what follows we consider the problem of optimizing parameters $\{\vartheta^k_{ij}\}$, which is equivalent to optimizing phase shift variables $\{v_n^k\}$.

\begin{Proposition} \label{Pro2} Suppose that $L^k_1=L^k_2=1$. For a given $k$, under the limit of large $N_t$ and $N_r$, the optimal values of phase shift variable $v_n^k, \;\; n=1,2,\ldots, N,$ are given by
\be \label{vvv} v_n^k=\pi(n-1)(\sin(\phi^{k1}_1)-\sin(\theta_2^{k1})).     \ee
\end{Proposition}
{\em Proof:} In order to maximize the system's achievable rate, all of the absolute values of the complex gains $\{|\vartheta^k_{11}|^2\}$ should be maximized. This implies that all of the absolute values of the corresponding factors $\{\xi^k_{11}=\ma_2^H(\theta_2^{k1})\mV^k\ma_1(\phi_1^{k1})\}$ should be maximized. Based on this argument, the desired result in \eqref{vvv} can be readily established. \hfill $\square$

\begin{Remark} \label{Re2} Under the IRS channels with pure LOS propagation,  Proposition 2 indicates that by employing the structure of ULA at the IRS, when $N_t$ and $N_r$ are very large, the optimized system controlling of the phase shift variables becomes easy, i.e., the IRS controller only requires to know these angles of arrival and departure related to the IRS. By the optimal control of phase shifts in (\ref{vvv}), we can have that $|\xi^k_{11}|^2=1$. However, without control of phase shifts, due to the fact that $\mV^k=\mI_N$ in this case, we can show in a similar fashion as in \cite{Yue1} that $|\xi^k_{11}|^2$ will tend to zero when $N$ grows without bound.
\end{Remark}

The IRS is composed of a large number of passive reflecting elements. This may lead to a prohibitive overhead of channel estimation in order to perform the optimal IRS control. For this reason, we assume that \be \label{vnk} v_n^k=\pi(n-1)\Delta^k_n, \;\; \Delta^k_n=\Delta^k,\;\; n=1,2,\ldots, N.     \ee
Based on this assumption, we can conclude that for any $i, j, k$,  $|\xi^k_{ij}|^2=1$ when $\Delta^k=\sin(\phi^{kj}_1)-\sin(\theta_2^{ki})$,  while $\lim_{N \to \infty}|\xi^k_{ij}|^2=0$ when $\Delta^k\neq \sin(\phi^{kj}_1)-\sin(\theta_2^{ki})$. Thus we naturally have an optimization strategy such that for a given $k$, we choose the strongest among all propagation paths to transmit and use its angles of arrival and departure related to the IRS to control the phase shifts. More precisely, we find the optimal index $i_0,j_0$ satisfying
\be \label{ooo}\omega(i_0,j_0)= \max_{i,j}\left\{\omega(i,j)=\frac{|\vartheta^k_{ij}|^2}{|\xi^k_{ij}|^2} \right\}.\ee

\begin{figure}[htb!]
\centering
\includegraphics[width=3.5in]{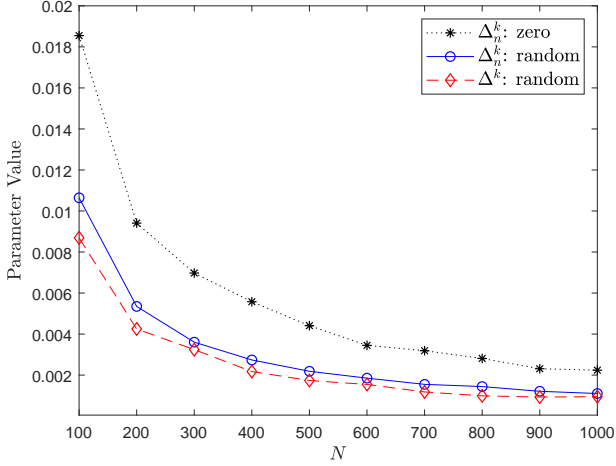}
\caption{Values of the parameter $|\xi^k_{ij}|^2$ versus $N$.}
\label{SIM2}
\end{figure}

For all other propagation paths with $i\neq i_0$ and $j \neq j_0$, they should not be useful when $N$ is very large and the total transmit power $P$ is limited. This is illustrated in Fig. \ref{SIM2}, which plots three curves that respectively correspond to three cases: (a) $\Delta^k_n$ is zero; (b) $\Delta^k_n$ is random; (c) only $\Delta^k$ is random.

The following Proposition summarizes the above analysis and results.

\begin{Proposition} \label{Pro3} When $N$,  $N_t$,  and $N_r$ are very large, but $P$ is limited,  the number of useful transmit paths provided by the IRS is $K$.
\end{Proposition}

\begin{Remark} \label{Re3} With the help of (\ref{RRR}) and (\ref{var}) and through the high SNR analysis,  by applying the notions of multiplexing gain and power (array) gain \cite{Tse}, we can further conclude that the IRS-assisted doubly-massive MIMO system's multiplexing gain is equal to $L_K$ while the power gain of each link related to the IRS is $N_rN_tN^2$. However, Proposition \ref{Pro3} implies that the number of truly useful transmission links provided by the IRS is only equal to $K$, provided that $N$ is very large. On the other hand, the contributions of the IRS, the transmitter, and the receiver to the total power gain of the IRS links are $N^2$, $N_t$, and $N_r$, respectively.
\end{Remark}

\section{The Scenario without Direct Signal Paths}

When $|\nu_l|^2, \;\; l=1,2,\ldots, L$ are too small, the $L$ direct signal paths between the transmitter and the receiver should be neglected. In this section, we perform analysis for this special case with pure LOS propagation and present a practical transmission scheme.

Suppose that $N_t$ and $N_r$ are very large. Then channel estimation becomes simple and only the angles of arrival and departure $\{\theta_1^{k1}, \theta_2^{k1}, \phi_1^{k1}, \phi_2^{k1}\}$ are required. Once these angles are determined, the optimal $\mathcal{V}$, $\mW_r$, and $\mW_r$ can be computed. Regarding power allocation, we may simply employ EPA, namely, $p_l=\frac{P}{K}, \;\;, l=1, 2, \ldots, K$. Thus, the achievable rate with EPA is given by
\be  \label{EEE} R_{\mathrm{EPA}}=\sum_{k=1}^{K}\log_2\lf(1+\frac{PN_rN_tN^2}{Kg_1^kg_2^k\sigma^2_n}\ri). \ee
If the destination (receiver) requires a particular date rate,  how much is the transmit power needed?  We consider this problem and obtain the following proposition.

\begin{Proposition} \label{Pro4} When $N_t$ and $N_r$ are very large,  the following average transmit power can ensure that the destination always achieves a given data rate $\bar{R}$.
\be \label{PPP} \bar{P}_{\mathrm{EPA}}=\frac{K\sigma_n^2\bar{g}_{\mathrm{I}}}{N_rN_tN^2}\cdot\left(2^{\bar{R}/K}-1\right), \ee
where
\be    \bar{g}_{\mathrm{I}}\;[\mathrm{dB}]=\sum_{i=1}^2 a_i+b_i\log_{10}(d_i)+c_i,\ee
and
\be  c_i=\frac{\ln 10}{20} \cdot 10^{\frac{\sigma_i [\mathrm{dB}]}{5}}. \ee
\end{Proposition}

{\em Proof:} For the achievable rate with EPA given in (\ref{EEE}), the following lower bound of $R_{\mathrm{EPA}}$ can be obtained by applying the arithmetic-geometric inequality:
\be \label{UUU} R_{\mathrm{EPA}}\geq K\log_2\lf(1+\frac{PN_rN_tN^2}{\sigma^2_n\sum_{k=1}^{K}(g_1^kg_2^k)}\ri). \ee
To satisfy the requirement of a given date rate $\bar{R}$ at the destination, the transmit power $P$ needs to satisfy:
\be   \log_2\lf(1+\frac{PN_r N_tN^2}{\sigma^2_n\sum_{k=1}^{K}g_1^kg_2^k}\ri) \geq \frac{\bar{R}}{K}. \ee
It follows that the average transit power is
\bee \bar{P}_{\mathrm{EPA}}&=&\mathbb{E}\left\{\frac{(2^{\bar{R}/K}-1)}{N_rN_tN^2}\lf(\sigma_n^2\sum_{k=1}^{K}g_1^kg_2^k\ri)\right\} \nnb \\
&=&\frac{\sigma_n^2(2^{\bar{R}/K}-1)}{N_r N_tN^2}\mathbb{E}\lf\{\sum_{k=1}^{K}g_1^kg_2^k\ri\}\nnb \\
&=&\frac{\sigma_n^2(2^{\bar{R}/K}-1)}{N_r N_tN^2}\sum_{k=1}^{K}\mathbb{E}(g_1^k)\mathbb{E}(g_2^k).\eee
Since
\bee \mathbb{E}(g_i^k)\;[\mathrm{dB}]&= &a_i+b_i1\log_{10}(d_i)+\mathbb{E}\chi((g_i^k))\nnb \\
&=&a_i+b_i\log_{10}(d_i)+c_i    \eee
and
\be \label{ccc} c_i=\mathbb{E}\chi((g_i^k))=\frac{\ln 10}{20} \cdot 10^{\frac{\sigma_i [\mathrm{dB}]}{5}}, \;\; i=1, 2,     \ee
the desired result (\ref{PPP}) can be easily obtained. 
\hfill $\square$

\begin{Remark} \label{Re4} Let $M=KN$ denote the total number of passive reflecting elements. It is known that the maximum multiplexing gain provided by the system is equal to $M$ if $K=M$ and $N=1$, whereas the maximum power gain provided by the system is equal to $N_tN_rM^2$ if $K=1$ and $N=M$. In practice, we can choose a solution that offers a tradeoff between power and multiplexing gains according to the propagation environments and requirements of the quality of service. Furthermore, through simple optimization based on (\ref{PPP}), it is readily derived that for a fixed $M$, the optimal $K$ needs to satisfy
\be  \bar{R}=3K (1-2^{-\bar{R}/K}).    \ee
When $\bar{R}$ is large, then the optimal $K$ is equal to or close to $\bar{R}/3$.
\end{Remark}

From a practical point of view, a wireless system needs to be designed to meet the requirements of both efficiency and reliability. In what follows, we consider a power allocation policy that is adaptive and can minimize the total transmit power under constraints on both the bit error probability and the transmission rate. Specifically, for the $k$th established data stream, its requirement in terms of the quality of service can be described by a required SNR at the receiver, denoted  $\overline{\mathrm{SNR}}_k$. From (\ref{UUU}), the instantaneous power allocated to the data stream is at least
\be P_k=\frac{\sigma_n^2g_1^kg_2^k\cdot\overline{\mathrm{SNR}}_k}{N_r N_t N^2}. \ee
Under the adaptive power allocation (APA) policy, a similar derivation yields the following minimum total transmit power:
\be \bar{P}_{\mathrm{APA}}=\mathbb{E}\lf\{\sum_{k=1}^{K}\frac{\sigma_n^2g_1^kg_2^k(2^{\bar{R}_k}-1)}{N_r N_tN^2}\ri\}=\frac{\sigma_n^2\bar{g}_{\mathrm{I}}}{N_rN_tN^2}\sum_{k=1}^{K}\overline{\mathrm{SNR}}_k. \ee

\section{Numerical Results}
In this section, we present numerical results to observe the performance behaviors of the considered IRS-assisted mmWave system and corroborate our analytial results.

\begin{figure}[htb!]
\centering
\includegraphics[width=3.54in]{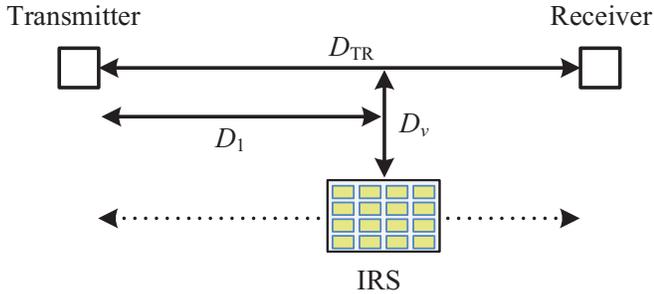}
\caption{Simulation setup.}
\label{SIM3}
\end{figure}

\begin{figure}[htb!]
\centering
\includegraphics[width=3.5in]{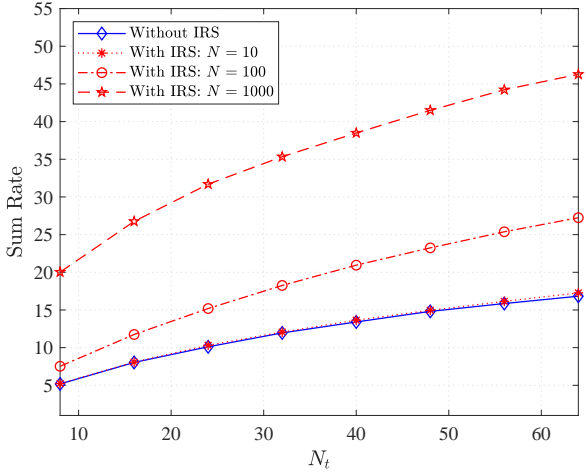}
\caption{Achievable rate versus $N_t$ for different values of $N$.}
\label{SIM6}
\end{figure}

For path-loss related parameters, we consider a setup where the IRS lies on a horizontal line which is in parallel to the line that connects the transmitter and the receiver. The distance between the transmitter and the the receiver is set to $D_{\mathrm{TR}} = 51$ meters and the vertical distance between two lines is set to $D_v = 2$ meters, as shown in Fig. \ref{SIM3}. Let $D_1$ denote the horizontal distance between the transmitter and the IRS. The transmitter-IRS distance and the IRS-receiver distance can then be calculated, respectively, as $D_{\mathrm{TI}} =\sqrt{D^2_1+D_v^2} $ and  $D_{\mathrm{IR}} =\sqrt{(D_{\mathrm{TR}}-D_1)^2+D_v^2}$. This means that $d_1=D_{\mathrm{TI}}$ and $d_2=D_{\mathrm{IR}}$. For the large-scale fading parameter in the NLOS channel between the transmitter and the receiver, the values of $a$, $b$ and $\sigma^2$ are set to be $a_0=72$, $b_0=29.2$, and $\sigma_0=8.7\;\mathrm{dB}$, as suggested in \cite{Akdeniz}. For the large-scale fading parameters in the LOS channels between the transmitter and the IRS, and between the IRS and the receiver, the values of $a$, $b$ and $\sigma^2$ are set to be $a_1=a_2=61.4$, $b_1=b_2=20$, and $\sigma_1=\sigma_2=5.8\;\mathrm{dB}$, also as suggested in \cite{Akdeniz}. We always fix $L_1^k=L_2^k=L=3$. Other parameters are set as follows: $P=30\;\mathrm{dBm}$ and $\sigma_n^2=-85 \;\mathrm{dBm}$ \cite{Wang1}. Except for Fig. \ref{SIM6}, each of the phase shifts, $v_n^k$, is determined according to (\ref{vnk}) and (\ref{ooo}).

\begin{figure}[htb!]
\centering
\includegraphics[width=3.5in]{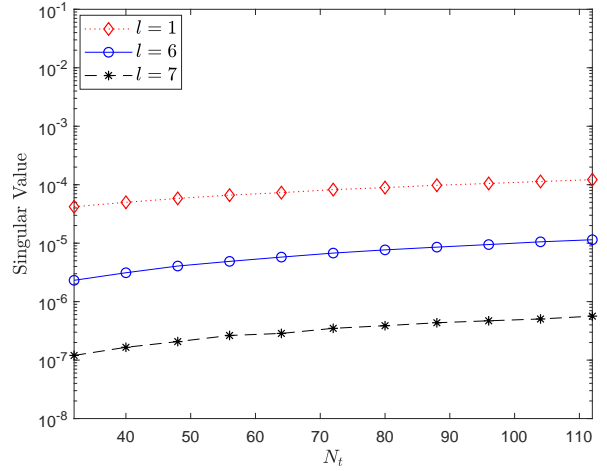}
\caption{Behavior of singular values of the channel matrix $\mathbb{H}$.}
\label{SIM4}
\end{figure}

First, the behavior of singular values of channel matrix $\mathbb{H}$ are studied. Let $\eta_1^k=\eta_2^k=\eta=5 \;\mathrm{dB}$, $K=3$, $N=300$, and $D=15$. It is expected that when $N_t$ and $N_r$ are large enough, the number of favorable and useful propagation paths for the examined case should be equal to $L_u=L+K=6$, as suggested by Proposition \ref{Pro3}. To confirm this, Fig. \ref{SIM4} plots the 1st, the 6th and 7th singular values (i.e., $l=1, 6, 7$), when $N_r$ increases from $32$ to $82$ as $N_t$ increases from $32$ to $112$. It can be seen from this figure that as $N_t$ increases, all singular values slowly increase, but the difference at $N_t=8$ and $N_t=64$ is small. The 7th singular value is very much smaller than the 6th singular value. Thus this figure verifies the conclusion stated in Proposition \ref{Pro3}.

\begin{figure}[htb!]
\centering
\includegraphics[width=3.5in]{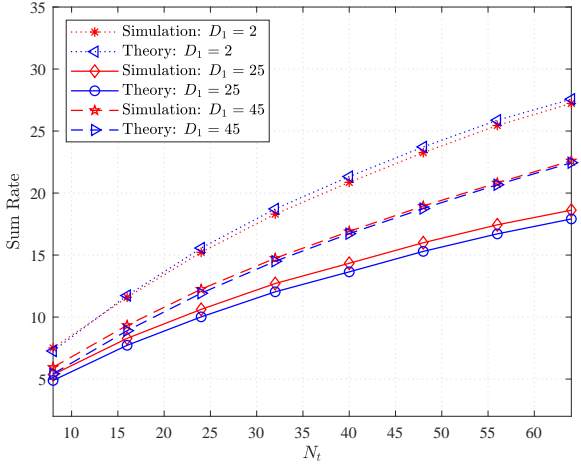}
\caption{Achievable rate versus $N_t$ for different values of $D_1$.}
\label{SIM5}
\end{figure}

Next, the achievable sum rate and its limit are examined. Now we set $\eta_1^k=\eta_2^k=5 \;\mathrm{dB}$, $\eta=-5 \;\mathrm{dB}$,  $K=3$ and $N=100$. When $N_r$ increases from $8$ to $36$ and $N_t$ increases from $8$ to $64$, Fig. \ref{SIM5} plots analytical results based on (\ref{RRR}) and Monte Carlo simulation results for different values of distances, namely, $D_1=2, 25, 45$ . It can be seen from this figure that when the IRS is close to the transmitter or the receiver, the system has better rate performance. Moreover, it can be observed that all of the simulation results are close to the analytical results as expected, which corroborates Proposition \ref{Pro1}.

In order to observe the rate performance improvement with increasing $N$ when $N_r$ increases from $8$ to $36$ and $N_t$ increases from $8$ to $64$, Fig. \ref{SIM6} plots the achievable sum rate for three different values of $N$, namely, $N=10, 100, 1000$. For this figure, we set $\eta_1^k=\eta_2^k=5 \;\mathrm{dB}$, $\eta=-5 \;\mathrm{dB}$, $K=3$ and $D_1=2$. For comparison, also plotted in the figure is the achievable sum rate when the system does not include the IRS. When $N=10$, the propagation paths created via the IRS are too weak and cannot be used in transmission. When $N=100$, however, the propagation paths created via the IRS become feasible and can be used to transmit data streams. Thus, the rate performance can be effectively improved. Furthermore, if $N$ is increased to $1000$,  the propagation paths created via the IRS become favorable and the sum rate with the IRS is about two times higher than that without the IRS when $N_t=64$ and $N_r=36$.

\begin{figure}[htb!]
\centering
\includegraphics[width=3.5in]{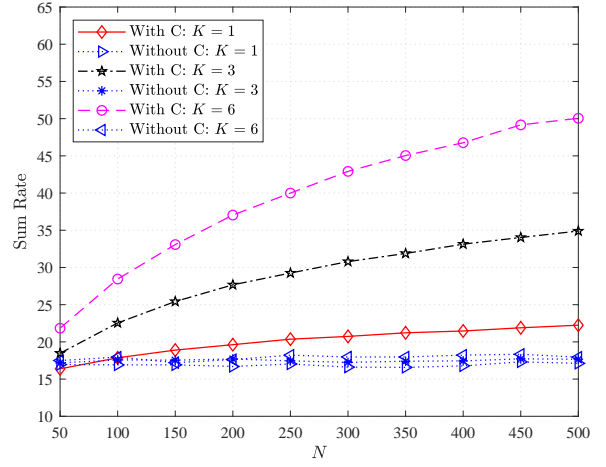}
\caption{Achievable rate versus $N$ for different values of $K$.}
\label{SIM7}
\end{figure}

\begin{figure}[htb!]
\centering
\includegraphics[width=3.5in]{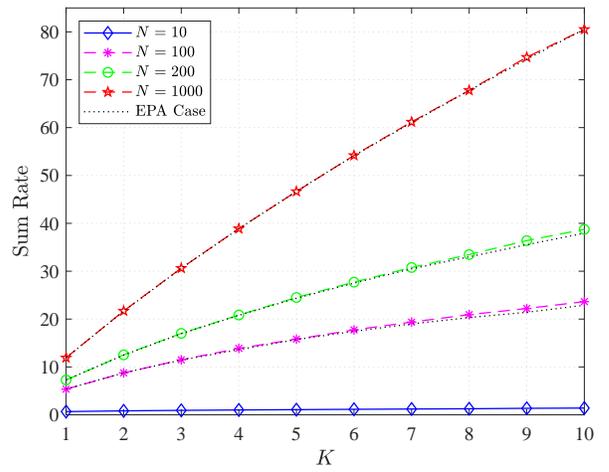}
\caption{Achievable rate versus $K$ for different values of $N$.}
\label{SIM8}
\end{figure}

Now we observe performance changes as $K$ increases when $\eta_1^k=\eta_2^k=5 \;\mathrm{dB}$, $\eta=-5 \;\mathrm{dB}$, $D_1=5$, $N_t=64$, and $N_r=36$. When $N$ increases from $50$ to $500$, Fig. \ref{SIM7} plots the achievable sum rate for three different values of $K$, namely, $K=1, 3, 6$. As expected, the achievable sum rate is significantly improved with increasing $K$. Moreover, Fig. \ref{SIM7} plots the three rate curves which correspond to the scenarios without control processing of the phase shifts. From this figure, the three curves are close to each other. This observation is expected and in agreement with Remark \ref{Re2}.

Under the case with $D_1=25$, $N_t=100$, and $N_r=100$, we consider the rate performance in the scenario where the direct signal paths are too weak and thus can be neglected while the IRS signal paths are LOS dominated. When $K$ increases from $1$ to $10$, Fig. \ref{SIM8} plots the achievable sum rate for $N=10, 100, 200, 1000$. For comparison, Fig. \ref{SIM8} also plots the achievable sum rate with EPA. It is interesting to see that all of the rate curves with EPA are quite close to the corresponding curves with the optimal power allocation (OPA). As expected, the rate performance with OPA and EPA is clearly improved as $N$ or $K$ increases, except for $N=10$. In the case with $N=10$, the rate is close to zero due to the fact that the propagation paths created via the IRS are too weak. For $M=KN=1000$, it can be determined by the numerical results that the solution of $(K=5, N=200)$ is a good option for the system design when comparing to the two extreme cases of $(K=1, N=1000)$ and $(K=10, N=100)$.

\begin{figure}[htb!]
\centering
\includegraphics[width=3.5in]{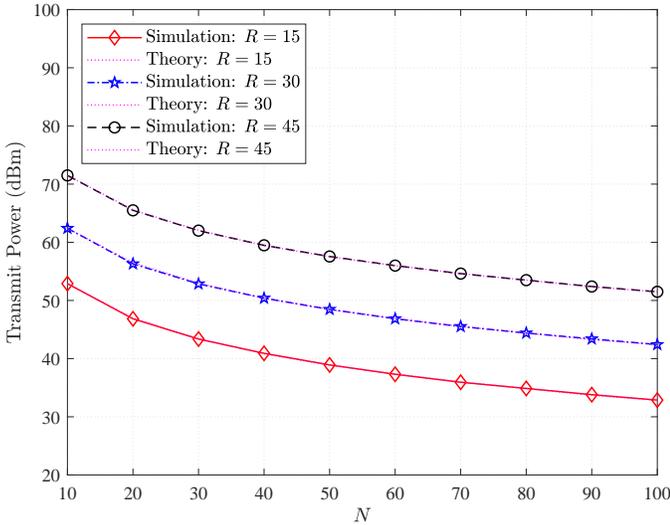}
\caption{Achievable rate versus $K$ for different values of $N$.}
\label{SIM9}
\end{figure}

Finally, we consider to verify Proposition \ref{Pro4} for $K=5$ under the condition where other parameters are set as the same as in Fig. \ref{SIM8}.  For given $\bar{R}=15, 30, 45$, we apply (\ref{UUU}) to compute the needed instantaneous transmit powers and then use them to conduct transmission with EPA. Fig.~\ref{SIM9} plots the average needed transmit powers when $N$ increases from $10$ to $100$.  Our simulation results shows that the obtained achievable rates are a little higher than the three corresponding required rates, and the simulation results with the transmit powers are almost the same as the analytical results given in (\ref{PPP}). This corroborates Proposition \ref{Pro4}. In addition, it can be seen from this figure that as $N$ is increased or $\bar{R}$ is decreased, the total required transmit power is decreased. In fact,  the transmit power is decreased linearly with the decreasing of $1/N^2$. Therefore, the difference of the required transmit powers of $N=10$ and $100$ for a given $\bar{R}$ should be $20 \;\mathrm{dB}$, which is verified by the simulation results in  Fig.~\ref{SIM9}.

\section{Conclusion}
IRS is new technology and envisioned to be a promising solution for the future 6G networks, but remains largely unexplored. This paper has investigated a point-to-point IRS-assisted mmWave doubly-massive MIMO system and derived expressions of the asymptotic sum rate when the numbers of antennas at the transmitter and receiver go to infinity. As a major difference compared to existing analysis and results, such as in \cite{Wang1} and \cite{Wu2}, it is shown in this paper that the optimal solutions of power allocation, precoding/combining, and IRS's phase shifts for the doubly-massive MIMO system can be realized independently without the need of joint processing. Such solutions are very convenient for the system design. In the future, we shall extend our analysis to the point-to-multiple points scenario.

\balance

\end{document}